\documentclass[a4paper,11pt]{article}

\pdfoutput=1
\usepackage{hyperref}
\usepackage{jheppub}

\usepackage{pifont}
\usepackage[numbers,compress]{natbib}
\usepackage{amsmath}
\usepackage{graphicx}
\usepackage{enumerate}
\usepackage{indentfirst}
\usepackage{subfig}
\usepackage{dsfont}
\usepackage{amssymb}
\usepackage{bm}
\usepackage{cases}
\title{\boldmath Bound states in the continuum are universal under the effect of minimal length}

\author[a,b]{Zhang Xiao}
\author[c,1]{Yang Bo,\note{Corresponding author.}}
\author[d]{Wei Chaozhen}
\author[a,b,e]{Luo Maokang}

\affiliation[a]{Department of Mathematics, Sichuan University, Chengdu, Sichuan, China}
\affiliation[b]{Nonlinear and Uncertain Engineering System Control Key Laboratory of Sichuan Province, Sichuan University, Chengdu, Sichuan, China}
\affiliation[c]{Department of Mathematics, Yunnan Normal University, Kunming, Yunnan, China}
\affiliation[d]{Department of Mathematical Sciences, Worcester Polytechnic Institute, Worcester, MA, USA}
\affiliation[e]{Department of Aeronautics and Astronautics, Sichuan University, Chengdu, Sichuan, China}

\emailAdd{zhangxiaomath@scu.edu.cn}
\emailAdd{boyang@ynnu.edu.cn}
\emailAdd{cwei4@wpi.edu}
\emailAdd{makaluo@scu.edu.cn}

\abstract{Bound states in the continuum (BICs) are generally considered unusual phenomena. In this work, we provide a method to analyze the spatial structure of particle's bound states in the presence of a minimal length, which can be used to find BICs. It is shown that the BICs are universal phenomena under the effect of the minimal length. Several examples of typical potentials, i.e., infinite potential well, linear potential, harmonic oscillator, quantum bouncer and Coulomb potential, et al, are provided to show the BICs are universal. The wave functions and energy of the first three examples are provided. A condition is obtained to determine whether the BICs can be readily found in systems. Using the condition, we find that although the BICs are universal phenomena, they are often hardly found in many ordinary environments since the bound continuous states perturbed by the effect of the minimal length are too weak to observe. The results are consistent with the current experimental results on BICs. In addition, we reveal a mechanism of the BICs. The mechanism explains why current research shows the bound discrete states are typical, whereas BICs are always found in certain particular environments when the minimal length is not considered.}
\begin{document}
\maketitle
\section{Introduction}
Bound states in the continuum (BICs) are considered unusual phenomena in quantum systems, the states of which are spatially bounded but have continuous energies. The BICs are counterintuitive eigenmodes of systems. In 1929, BICs were first presented by von Neumann and Wigner \cite{1}; until recently, they were constructed as follows:
\begin{itemize}\setlength{\itemsep}{0pt}
  \item fine-tuning parameters in the Schr\"{o}dinger equation to build tailored potentials \cite{1,2,24};
  \item decoupling all the continuum states due to symmetry \cite{8,7};
  \item describing matter by the L\'{e}vy path integral, i.e., the fractional Schr\"{o}dinger equation \cite{3,4};
  \item using the direct and via-the-continuum interaction between particles \cite{6,10,25,26,27,28}.
\end{itemize}\par
These studies always show that the BICs are unusual phenomena that need to be constructed or realized in a particular manner \cite{1,2,24,8,7,3,4,6,10,25,26,27,28,5,9,18}, since the current results show that common quantum systems usually have bound discrete states and unbound continuous states in natural potentials \cite{15}. For example, 
Ref. \cite{30} demonstrated BICs in a one-dimensional quantum wire with two impurities induced due to Fano interference.
In Ref. \cite{25}, BICs were trapped or guided modes, with their frequencies in the frequency intervals of the radiation modes.
In Ref. \cite{26}, BICs were produced in a system of $n$ two-level quantum emitters coupled with a one-dimensional photon field.
In Ref. \cite{27}, BICs were considered in the nonrelativistic reduction of quasipotential equations in QED and Wick-Cutkosky models.
In Ref. \cite{28}, the formation of BICs was provided in a whispering gallery resonator coupled to a one-dimensional waveguide.
In particular, BICs were realized in the finite lattices with arbitrary boundaries \cite{65,66}. 
Subsequent experiments also provided the observation of BICs in certain environments, such as electronic BICs in semiconductor heterostructures \cite{6}, optical BICs in planar optical waveguide arrays \cite{7}, and robust and non-symmetrically protected BICs in the periodic extended structure \cite{5}. \par
However, in this paper, we find that BICs are universal phenomena under the effect of minimal length and can be found in most potentials, even in ordinary potentials with single particles such as the infinite potential well, quantum bouncer, linear potential, harmonic oscillator and Coulomb potential. The minimal length is a universal effect of quantum gravity, which is considered as quantum gravity corrections to all quantum Hamiltonian \cite{11}. It has long been known that string theory and various models of quantum gravity predict the existence of the minimal length \cite{4,11,12,20,21,22,23,32,35,48,49}. A simple and accepted way to introduce the minimal length is the generalized uncertainty principle (GUP) \cite{12}:
\begin{equation}\label{eq:GUP-relations}
\Delta X\Delta P\geq\frac{\hbar}{2}\{1+\beta[(\Delta P)^2+\langle P \rangle^2]\},
\end{equation}
where $X$ and $P$ are the quantities with the minimal length, $\langle P \rangle$ represents the average of $P$, and $\beta$ is a deforming parameter with upper bound of $10^{18}\thicksim10^{78}$ in different physical frameworks \cite{11,22,23}. It is a modification of the Heisenberg uncertainty principle (HUP); when $\beta=0$, Eq.\eqref{eq:GUP-relations} recovers the HUP. \par
Thus, the study of quantum theories characterized by the minimal length, therein involving high-energy physics, cosmology and black holes, has become an active area of research, especially in the quantum regime. For example, in Ref. \cite{62}, authors investigated the impact of the GUP. In Ref. \cite{32}, the effects of the GUP on the classical and quantum cosmology of a closed Friedmann universe were studied. In Ref. \cite{63}, the GUP that leads to vanishing quantum effect was studied. In Ref. \cite{61}, a new higher order GUP was presented. In Ref. \cite{62}, the ground state energy of the hydrogen atom was studied in the presence of the minimal length. In Ref. \cite{51,52}, the inflationary predictions for the cosmic microwave background were studied based under the minimal length. In Ref. \cite{35}, the authors showed the GUP results for minimum uncertainty wave packets. In Refs. \cite{22,23}, the upper bound of the deforming parameter $\beta$ of the minimal length was discussed. In Ref. \cite{64}, the authors extended the GUP discloses a self-complete characteristic of gravity in order to overcome some current limitations to the framework.\par

In this paper, we suggest a method to analyze the spatial structure of a particle's bound states under the effect of the minimal length, which can be used to find BICs in potentials. Existing methods to this end without a minimal length are not suitable \cite{13,14,15}. Using this approach, we show that the BICs are universal phenomena under the effect of the minimal length. Several examples with ordinary potentials, i.e., the infinite potential well, linear potential, harmonic oscillator, Poschl-Teller potential, quantum bouncer, half oscillator, quantum bouncer in a closed court, harmonic oscillator plus Dirac delta function and Coulomb potential, are provided to show that the BICs are universal. The specific wave functions and energy of the first three examples are provided. A condition is provided to determine whether the BICs can be readily found in the systems: (1) when the deforming parameter $\beta$ of the minimal length satisfies that $\hbar\beta[(\Delta P)^2+\langle P\rangle ^2]$ is close to the Planck constant $\hbar$, the BICs can be easily found, and (2) when $\beta$ satisfies $\beta[(\Delta P)^2+\langle P\rangle ^2]\ll1$, the BICs are very inconspicuous. This condition may help researchers to set suitable environmental variables of potential to obtain obvious BICs. Three examples are discussed.\par
In addition, we reveal a mechanism of the BICs. The mechanism demonstrate that why the current research shows that bound discrete states are universal, whereas the BICs are always found in certain particular environments when the minimal length is not considered. The minimal length can influence quantum systems to produce extra states. These states have an energy found in the energy gaps between the standard discrete energy levels to make the energy continuous, namely, the effect of the minimal length allows particles to not be restricted to move in some discrete energy levels. These cause the BICs to be universal phenomena under the effect of the minimal length. Then, when $\beta$ satisfies $\beta[(\Delta P)^2+\langle P\rangle ^2]\ll1$, the influence of the minimal length becomes negligible. All these extra states caused by the minimal length become too weak to be readily found; thus, we often assume that particles can not move at these energy, i.e., the energy gaps are formed. The continuous energy becomes discrete, which causes the BICs to always not be typical in current results when the effect of the minimal length is considered. Our conclusions are consistent with the current experimental results on BICs.\par
This work is organized as follows. In Sec. \ref{sec:GM}, we build a method to find BICs in potentials under the effect of the minimal length and provide a condition to determine whether the BICs can be readily observed in systems. Then, we find that BICs are universal and find a mechanism of BICs in the presence of the minimal length. In Sec. \ref{sec:E}, as examples, we find BICs in three ordinary potentials: the infinite potential well, linear potential and harmonic oscillator. We find the condition that makes BICs could be observed. The specific wave functions, degeneracy and energy of the three examples are provided.
\section{General method}\label{sec:GM}
In this section, we obtain two main results. \par
First, we find that the space of the solutions of the Schr\"{o}dinger equation with minimal length has $4$ degrees of freedom (DFs), which is greater than that of the standard Schr\"{o}dinger equation of $2$, by expressing the basis of the space of the solutions.
The extra DF provided by the minimum length allows the particles to not be restricted to move at certain energy levels, causing a continuous energy. \par
Thus, we divide bound states into three cases and provide a way to find BICs by analyzing the DFs of systems that are restricted by boundary conditions to reduce the DFs. Using this approach, we find that BICs exist in these cases. If $\beta=0$, namely, the system is not considered to be influenced by the minimal length, the systems have $0$ DFs; thus, BICs do not exist. This shows one of the main results in this paper, i.e., BICs are universal under the effect of the minimal length because of the sufficient extra DFs. Many ordinary potentials, such as the infinite potential well, linear potential, harmonic oscillator, Poschl-Teller potential, quantum bouncer, half oscillator, quantum bouncer in a closed court, harmonic oscillator plus Dirac delta function and Coulomb potential, satisfy the conditions of these three cases.\par
Second, since the effect of the minimal length causes the extra DFs and hence causes the BICs, we further analyze the effect of the minimal length on quantum systems. We find that the extra DFs are determined by the value of $\hbar\beta[(\Delta P)^2+\langle P\rangle ^2]$ which is determined by the environmental variables of potentials. Thus, we obtain another main result: a condition to determine whether the BICs can be readily found in the systems. \par
Based on these two results, we obtain a mechanism for BICs under a minimal length.

\subsection{BICs are universal under the effect of a minimal length}\label{sec:B-U}
We consider a particle moving in a time-independent potential $V(X)$. The quantities $X$ and $P$ with minimal length can be represented as $X=x,P=p(1+\beta' {p}^2)$ \cite{11}, where $x$ and $p$ are the standard quantities without the minimal length and $\beta'=\beta/3$. Substituting these into the Hamiltonian
\begin{equation*}
H=\frac{P^2}{2m}+V(X),
\end{equation*}
and using the correspondence relation of the momentum $p\rightarrow -i\hbar(\mathrm{d}/\mathrm{d}x)$, the dynamical evolutions with the minimal length of the wave functions (the Schr\"{o}dinger equation with the minimal length) are given by
\begin{equation}\label{eq:S-minimal}
\frac{\beta'\hbar^4}{m}\frac{\mathrm{d}^4 \varphi(x)}{\mathrm{d} x^4}-\frac{\hbar^2}{2m}\frac{\mathrm{d}^2 \varphi(x)}{\mathrm{d} x^2}+[V(x)-E]\varphi(x)=0.
\end{equation}\par
There are two main kinds of methods to reduce the above fourth-order Schr\"{o}dinger equation to a second-order one at present:
\begin{enumerate}\setlength{\itemsep}{0pt}
  \item[(i)] omitting the terms with $\beta^2$ to reduce the order of the fourth-order Schr\"{o}dinger equation after using some substitution of wave functions \cite{50,54};
  \item[(ii)] studying the fourth-order Schr\"{o}dinger equation of some very particular potentials in the momentum representation (by employing Fourier transform) \cite{12,57}.
\end{enumerate}\par
However, using these methods, to some extent, only some particular solutions of the fourth-order Schr\"{o}dinger equation can possibly be obtained (see App. \ref{app} for more details) in some particular situations. Thus, we should be very careful with this method since it would influence the solution space of the equation and hence alter the properties of the solution states. We have shown its influence on the solutions in two different situations in App. \ref{app}. \par
Since there exists the parameter $\beta'$ in the highest order term $[\beta'\hbar^4\mathrm{d}^4 \varphi(x)]/(m\mathrm{d} x^4)$ of the above Schr\"{o}dinger equation Eq. \eqref{eq:S-minimal}, the equation is a singular perturbation system \cite{55,56}. If we truncate the term with $\beta'$ or set $\beta'=0$ in Eq.~\ref{eq:S-minimal}, the fourth-order equation is reduced to a second-order one and this may change the space solution and alter the properties of the solution states. This fourth-order term is essential to the equations and cannot be simply omitted in some particular situations, because it not only accounts for the effect of the quantum-gravitational fluctuations of the background metric \cite{53}, but also determines the solution space of the equation \cite{55,56}. \par
This kind of system is sensitive to its the highest derivative terms; transforming it may cause exotic phase transitions in its solution space \cite{55,56}. This causes exotic results when we study the system using the above two methods in some situations: for example, only some particular solutions are obtained; the states caused by the singular perturbation are ignored; complete wave functions can not be obtained; or certain phenomena are ignored \cite{58,59}(see App. \ref{app} for more details). In particular, Ref. \cite{4} showed that we can only obtain part of the solutions of the Schr\"{o}dinger equation of the minimal length in the linear potential in the momentum representation (when the parameter $\alpha=1$ of the Schr\"{o}dinger equation in Ref. \cite{4} is Eq. \eqref{eq:S-minimal}). Therefore, we should be very careful with this method since it would influence the solution space of the equation and hence alter the properties of the solution states. (For the linear potential and quantum harmonic oscillator, if we transform the time-independent Schr\"{o}dinger equation in the minimal length from the position representation to the momentum representation, the equations become regular perturbation systems from the singular perturbation systems, because not all the highest order derivative terms of the two equations include the parameters $\beta$ \cite{55,56}). In this article, we provide a method to directly analyze the fourth-order Schr\"{o}dinger equation in the position representation, and the results indicate some counterintuitive conclusions.\par
Next, we will address the method to analyze the spatial structure of a particle's bound states and find BICs in most ordinary potentials with a single particle under the effect of the minimal length, more precisely obtaining the energy and wave functions of the systems. For simplicity, we rewrite Eq. \eqref{eq:S-minimal} in matrix form:
\begin{equation}\label{eq:S-minimal-matrix}
\frac{\mathrm{d} \Phi(x)}{\mathrm{d} x}=A(x)\Phi(x),
\end{equation}
where
\begin{equation*}
A(x)=\left(
  \begin{array}{cccc}
    0 & 1 & 0 & 0 \\
    0 & 0 & 1 & 0 \\
    0 & 0 & 0 & 1 \\
    \frac{m[E-V(x)]}{\beta'\hbar^4} & 0 & \frac{1}{2\beta'\hbar^2} & 0 \\
  \end{array}
\right);~~
\Phi(x)=\left(
          \begin{array}{c}
            \varphi(x) \\
            \frac{\mathrm{d} \varphi(x)}{\mathrm{d} x} \\
            \frac{\mathrm{d}^2 \varphi(x)}{\mathrm{d} x^2} \\
            \frac{\mathrm{d}^3 \varphi(x)}{\mathrm{d} x^3} \\
          \end{array}
        \right).
\end{equation*}
$A(x)$ describes the Hamiltonian of the system, and $\Phi(x)$ represents the particle's motion. \par
First, we show that the space of the solutions of Eq. \eqref{eq:S-minimal} has $4$ DFs in the following mild mathematical derivations. This means that the general solutions of Eq. \eqref{eq:S-minimal} have $4$ free undetermined coefficients (that is, $C_1,\cdots,C_4$ in Eq. \eqref{eq:solution}). The $4$ DFs are illustrated in Fig. \ref{fig:1}(a) and (b) as $\omega_1,\cdots,\omega_4$, namely, any linear combinations of $\omega_1,\cdots,\omega_4$ are solutions of Eq. \eqref{eq:S-minimal}. However, the standard 1-D Schr\"{o}dinger equation has $2$ DFs at most, which are illustrated as $\omega_1$ and $\omega_2$ in Fig. \ref{fig:1}(c).\par
For any four linearly independent vectors $\varepsilon_i\in\mathds{R}^4$, where $i=1,\cdots,4$, and a point $x_0\in\mathds{R}$, there exist solutions $\Phi_1(x),\cdots,\Phi_4(x)$ of Eq. \eqref{eq:S-minimal-matrix} satisfying $\Phi_i(x_0)=\varepsilon_i$ \cite{19}. Let $\varphi_i(x)$ represents the first component of $\Phi_i(x)$; then, there exists a non-zero $x=x_0$ of the Wronskian determinant $W(\varphi_1(x),\varphi_2(x),\varphi_3(x),\varphi_4(x))$:
\begin{equation*}
W(\varphi_1(x),\varphi_2(x),\varphi_3(x),\varphi_4(x))|_{x=x_0}=\det(\Phi_1(x_0),\cdots,\Phi_4(x_0))=\det(\varepsilon_1,\cdots,\varepsilon_4)\neq0.
\end{equation*}
Thus, $\varphi_1(x),\cdots,\varphi_4(x)$ are linearly independent. Any solutions of Eq. \eqref{eq:S-minimal} can be represented by them:
\begin{equation}\label{eq:solution}
\varphi(x)=\sum\limits_{k=1}^4 C_k\varphi_k(x),
\end{equation}
where $C_1,\cdots,C_4$ are $4$ arbitrary coefficients. This shows that the space of the solutions of Eq. \eqref{eq:S-minimal} is $4$ dimensional, and $\varphi(x)$ has $4$ DFs.\par
Second, we propose a method to determine whether BICs exist in a system under the effect of the minimal length. Applying the method, we show that BICs can be found in many ordinary potentials. To build this method, we divide all the boundary conditions that limit the motion of particles into two types. One type can ensure states being bounded, such as $\lim\limits_{x\rightarrow\pm\infty}\varphi(x)=0$; we call them key boundary conditions (KBCs). The other type is an unnecessary condition for states being bounded, such as $\lim\limits_{x\rightarrow0}\varphi(x)=0$ in the Coulomb potential; we call them non-key boundary conditions (non-KBCs). Then, we divide the three cases up to discuss the energy and bound states, therein being one of the main results in this work.\par
\begin{figure}[!h]
\centering
\subfloat[Case I]{\includegraphics[width=3.5cm]{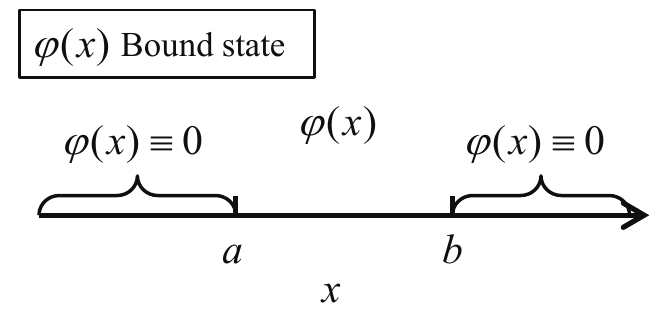}}
\subfloat[Case II]{\includegraphics[width=3.5cm]{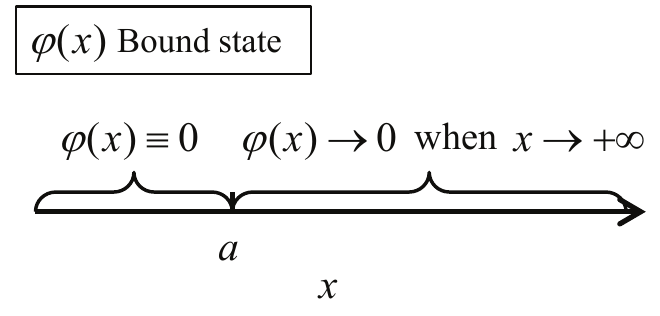}}
\subfloat[Case III]{\includegraphics[width=3.5cm]{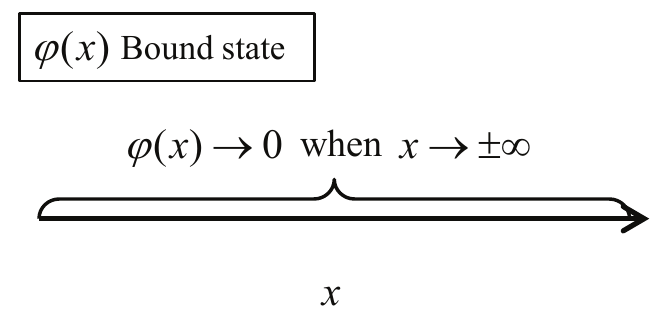}}
\caption{The conditions of bound state (KBCs) for different cases}\label{fig:3}
\end{figure}

Case I--The KBCs satisfy the following conditions: there exist two different $x_a,x_b(x_a<x_b)$ such that the wave function $\varphi(x)=0$ in the regions $(-\infty,x_a]$ and $[x_b,+\infty)$, as shown in Fig. \ref{fig:3}(a). If there are fewer than $2$ non-KBCs, combining the KBCs, there are $3$ conditions at most, which cannot determine all $4$ undetermined coefficients of Eq. \eqref{eq:solution}
\begin{equation*}
\varphi(x)=\sum\limits_{k=1}^4 C_k\varphi_k(x).
\end{equation*}
Thus, the energy $E$ is arbitrary. The KBCs ensure the states of the system to be bounded so that the BICs exist in these potential under the effect of the minimal length. Many potentials satisfy the conditions of Case I, e.g., the infinite potential well, asymmetric infinite well \cite{16}, Dirac delta function in the infinite square well, Poschl-Teller potential \cite{14} and quantum bouncer in a closed court \cite{16}.\par

Case II--The KBCs satisfy the conditions $\lim\limits_{x\rightarrow+\infty}\varphi(x)=0$ (or $\lim\limits_{x\rightarrow-\infty}\varphi(x)=0$), and there exists one point $x=x_a$ (or $x_b$) such that wave function $\varphi(x)=0$ in the region $(-\infty,x_a]$ (or $[x_b,+\infty)$), as shown in Fig. \ref{fig:3}(b).
This type of KBC would determine $3$ undetermined coefficients.

If there are no non-KBCs, $3$ of $4$ undetermined coefficients in Eq. \eqref{eq:solution} are determined. Thus, the energy $E$ is also arbitrary. Again, we can find the BICs in these potentials under the effect of the minimal length. Many potentials satisfy the above-mentioned conditions, e.g., linear potential, quantum bouncer \cite{16}, half oscillator \cite{16} and some other half infinite potential wells.\par
Case III--The KBCs satisfy the conditions $\lim\limits_{x\rightarrow\pm\infty}\varphi(x)=0$, as shown in Fig. \ref{fig:3}(c). They can only determine $2$ undetermined coefficients.

If there are fewer than $2$ non-KBCs, combining the KBCs, they can determine $3$ of $4$ coefficients in Eq. \eqref{eq:solution} at most. Thus, the energy $E$ is arbitrary. Again, we can find the BICs in these potentials under the effect of the minimal length. Many potentials, such as quantum harmonic oscillator, harmonic oscillator plus Dirac delta function \cite{16} and Coulomb potential, satisfy the mentioned boundary conditions.\par

As a result, we obtain BICs in these three cases. Most of potentials that have bound states belong to these three cases; thus, a counterintuitive conclusion is obtained: BICs are universal phenomena under the effect of a minimal length.
\subsection{Condition for determining whether the BICs can be readily observed in systems}\label{sec:C}
\begin{figure}[h]
\centering
\includegraphics[width=11cm]{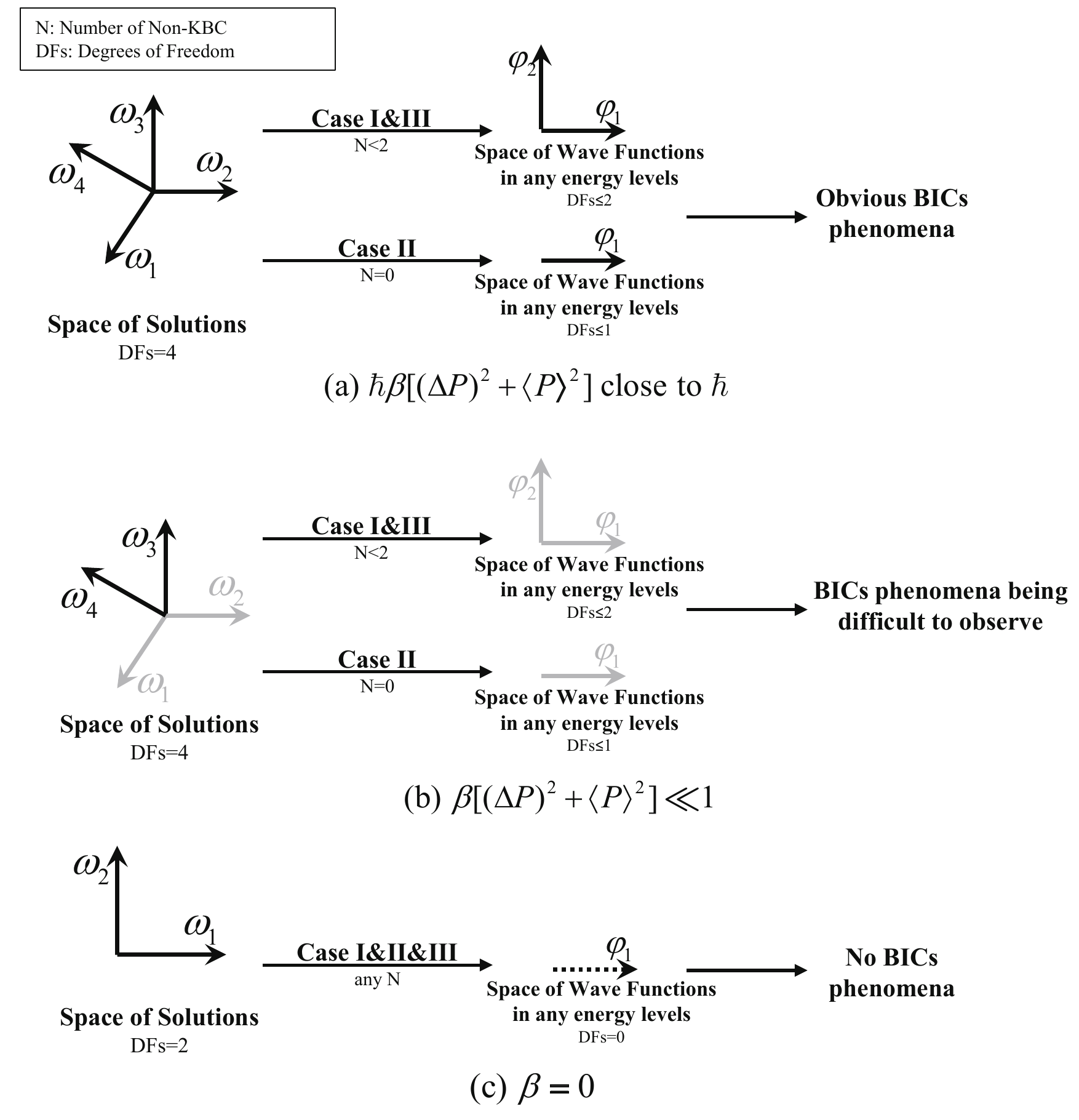}
\caption{The mechanism of the BICs under the effects of the minimal length and the conditions for determining whether the BICs can be readily found in the systems. $\omega_i$ are the bases of the solution space and $\varphi_i(x)$ are degenerate states. The dotted line of $\varphi_1$ represents a discrete energy, and the solid lines represent a continuous energy. The black lines represent energies that can be readily found, and the gray lines represent energies that are not readily found.}\label{fig:1}
\end{figure}

The time-independent Schr\"{o}dinger equation (Eq. \eqref{eq:S-minimal}) in the presence of the minimal length is
\begin{equation*}
\frac{\beta'\hbar^4}{m}\frac{\mathrm{d}^4 \varphi(x)}{\mathrm{d} x^4}-\frac{\hbar^2}{2m}\frac{\mathrm{d}^2 \varphi(x)}{\mathrm{d} x^2}+[V(x)-E]\varphi(x)=0.
\end{equation*}
Let $\eta^4=\beta'^{-1}$, $a=\eta^2/(4\hbar^2)$, and $b(x)=(V(x)-E)m/\hbar^4$; the time-independent Schr\"{o}dinger equation with the minimal length becomes
\begin{equation}\label{eq:S-minimal-G}
\frac{\mathrm{d}^4\varphi(x)}{\mathrm{d}x^4}-2\eta^2a\frac{\mathrm{d}^2\varphi(x)}{\mathrm{d}x^2}+b(x)\eta^4\varphi(x)=0.
\end{equation}
If the energy $E>0$, solving the above Eq. \eqref{eq:S-minimal-G}, we have the solutions
\begin{equation}\label{eq:S-minimal-G:s}
\varphi(x)=C_1\omega_1(x)+C_2\omega_2(x)+C_3\omega_3(x)+C_4\omega_4(x),
\end{equation}
where $C_1,\cdots,C_4$ are arbitrary coefficients, and
\begin{equation}\label{eq:omega-G}
\begin{split}
\omega_j(x)= & \frac{1}{\sqrt{\lambda_j(x)}\sqrt[4]{a^2-b(x)}}\cdot\\
&\exp[\eta\int_{x_0}^x \lambda_j(\chi)\mathrm{d}\chi-\frac{1}{2}\int_{x_0}^x\lambda_j'(\chi)\frac{1}{\sqrt{a^2-b(\chi)}}\mathrm{d}\chi][1+O(\eta^{-1})];j=1,2,3,4,
\end{split}
\end{equation}
where $x_0$ is an arbitrary point of the particle's position and
\begin{align}\label{eq:lambda-G-1234}
\lambda_1(x)&=\sqrt{a+\sqrt{a^2-b(x)}}; & \lambda_2(x)&=-\sqrt{a+\sqrt{a^2-b(x)}};\\ 
\lambda_3(x)&=\sqrt{a-\sqrt{a^2-b(x)}}; & \lambda_4(x)&=-\sqrt{a-\sqrt{a^2-b(x)}}.
\end{align}
$\omega_1(x),\omega_2(x),\cdots,\omega_4(x)$ determine the solution space of the Schr\"{o}diner equation with the minimal length; thus, there are $4$ DFs, as shown in Fig. \ref{fig:1} (a) and (b). The standard Schr\"{o}diner equation, namely, when it is not perturbed by the effect of the minimal length, has $2$ DFs, as shown in Fig. \ref{fig:1} (c). From the results in Sec. \ref{sec:B-U}, the extra DFs caused by the effect of the minimal length lead to the energy of the bound states being continuous. Thus, the effect of the minimal length leads to the existence of BICs in universal potentials. When the minimal length has a prominent influence on the system, BICs are easily found; otherwise, when the influence of the minimal length is negligible, BICs can hardly be observed. Next, we provide the condition for determining whether BICs are easily found.\par
In the Schr\"{o}dinger equation, Eq. \eqref{eq:S-minimal}, the fourth-order item $(\beta'\hbar^4/m)[\mathrm{d}^4 \varphi(x)/\mathrm{d} x^4]$ is produced by the perturbation of the minimal length, which causes the extra DFs. By the GUP:
\begin{equation*}
\Delta X\Delta P\geq\frac{\hbar}{2}\{1+\beta[(\Delta P)^2+\langle P \rangle^2]\},
\end{equation*}
the fourth-order item is determined by the extra item $0.5*\hbar\beta[(\Delta P)^2+\langle P \rangle^2]$ in the GUP. Compared with the HUP $\Delta X\Delta P\geq0.5*\hbar$, if the magnitude of the extra item perturbed by the minimal length is close to the magnitude of the item determined by the Planck length, then
\begin{equation}
\hbar\beta[(\Delta P)^2+\langle P \rangle^2] \text{ is close to Planck constant }\hbar.
\end{equation}
The influence of the minimal length is prominent and provides extra DFs. Thus, the BICs are readily found, as shown in Fig. \ref{fig:1} (a).\par
If the magnitude of the item perturbed the minimal length is much less than the magnitude of the item determined by the Planck length, namely, satisfies that
\begin{equation}
\beta[(\Delta P)^2+\langle P \rangle^2] \ll 1.
\end{equation}
The influence of the minimal length is negligible, and the probabilities of two states $\omega_1,\omega_2$ that provide $2$ DFs are much smaller than the others probabilities of the superposition states. The system is close to the state with $2$ DFs (without the minimal length); thus, the BICs are very inconspicuous, that is, they are difficult to observe, as shown in Fig. \ref{fig:1} (b). The grey lines represent the states that are difficult to find at certain energy levels. However, since there are still $4$ DFs of the system, BICs still exist.\par
If the effect of the minimal length completely vanishes, that is, $\beta=0$, there is once again $2$ DFs of the system. Since there is no extra DF, the energy of the bound states is not free to take any value; thus, it is discrete, as shown in Fig. \ref{fig:1} (c).\par
As a result, BICs are universal under the effect of a minimal length; however, they may be difficult to observe due to the intensity of the perturbation caused by the minimal length. \par
Moreover, a mechanism of the BICs is provided. Figure \ref{fig:1} illustrates the mechanism of the BICs: the BICs are universal under the effect of a minimal length. The intensity of the effect of the minimal length determines whether the BICs can easily be found. When $\hbar\beta[(\Delta P)^2+\langle P\rangle ^2]$ is close to Planck constant $\hbar$, the system has $4$ DFs at most; thus, the particle cannot be restricted to move only at certain special energy levels, causing a continuous energy. When $\beta[(\Delta P)^2+\langle P\rangle ^2]\ll 1$, the system is close to the state with $2$ DFs (without the minimal length), causing some states (gray lines in Fig. \ref{fig:1}(b)) at certain energies to be difficult to observe. Although the energy is still continuous, it is difficult to find since these energy levels' states (gray lines in Fig. \ref{fig:1}(b)) are difficult to observe. Thus, it is often considered that particles cannot move at these energies, i.e., they have discrete energies; it is also often considered that the BICs do not exist without a minimal length. When the minimal length completely vanishes ($\beta=0$), there are $2$ DFs; there are insufficient DFs to produce extra states such that the energy is discrete, namely, no BICs exist. This is consistent with the current results: the bound discrete states are typical; however, the BICs are always found in certain particular environments when a minimal length vanishes. \par

\section{Examples}\label{sec:E}
In this section, we study an example of each case in Sec. \ref{sec:B-U} by the method built in Sec. \ref{sec:GM}. We find that BICs exist in these ordinary potentials with single particles under the effect of the minimal length: infinite potential well, linear potential, harmonic oscillator. We also provide some values set of the environmental variables to obtain obvious BICs in these ordinary potentials. The specific wave functions, degeneracy and energy of these examples are provided.
\subsection{Infinite potential well}
We consider a particle moving in the potential $V(x)$ satisfying $V(x)=0$ in the region $(-a,a)$ and $V(x)=+\infty$ in the other regions, where $a>0$. The time-independent Schr\"{o}dinger equation with minimal length in the region $(-a,a)$ is
\begin{equation}\label{eq:S-minimal-I}
\frac{\beta'\hbar^4}{m}\frac{\mathrm{d}^4 \varphi(x)}{\mathrm{d} x^4}-\frac{\hbar^2}{2m}\frac{\mathrm{d}^2 \varphi(x)}{\mathrm{d} x^2}-E\varphi(x)=0,
\end{equation}
where the boundary conditions are $\varphi(\pm a)=0$. If the energy $E>0$, solving the above Eq. \eqref{eq:S-minimal-I}, we have
\begin{equation}\label{eq:S-minimal-I:s}
\varphi(x)=C_1\exp(\lambda_1 x)+C_2\exp(\lambda_2 x)+C_3\cos(\lambda_3 x)+C_4\sin(\lambda_3 x),
\end{equation}
where $C_1,\cdots,C_4$ are arbitrary coefficients, and
\begin{align*}
\lambda_1&=\frac{1}{2\hbar}\sqrt{\frac{1+\sqrt{1+16E\beta'm}}{\beta'}}; & \lambda_2&=-\lambda_1; &
\lambda_3&=\frac{1}{2\hbar i}\sqrt{\frac{1-\sqrt{1+16E\beta'm}}{\beta'}}.
\end{align*}\par
By the method in Case I, the conditions $\varphi(\pm a)=0$ cannot determine all the coefficients of Eq. \eqref{eq:S-minimal-I:s}, and BICs exist. \par
\begin{figure}[h]
\centering
\includegraphics[width=11cm]{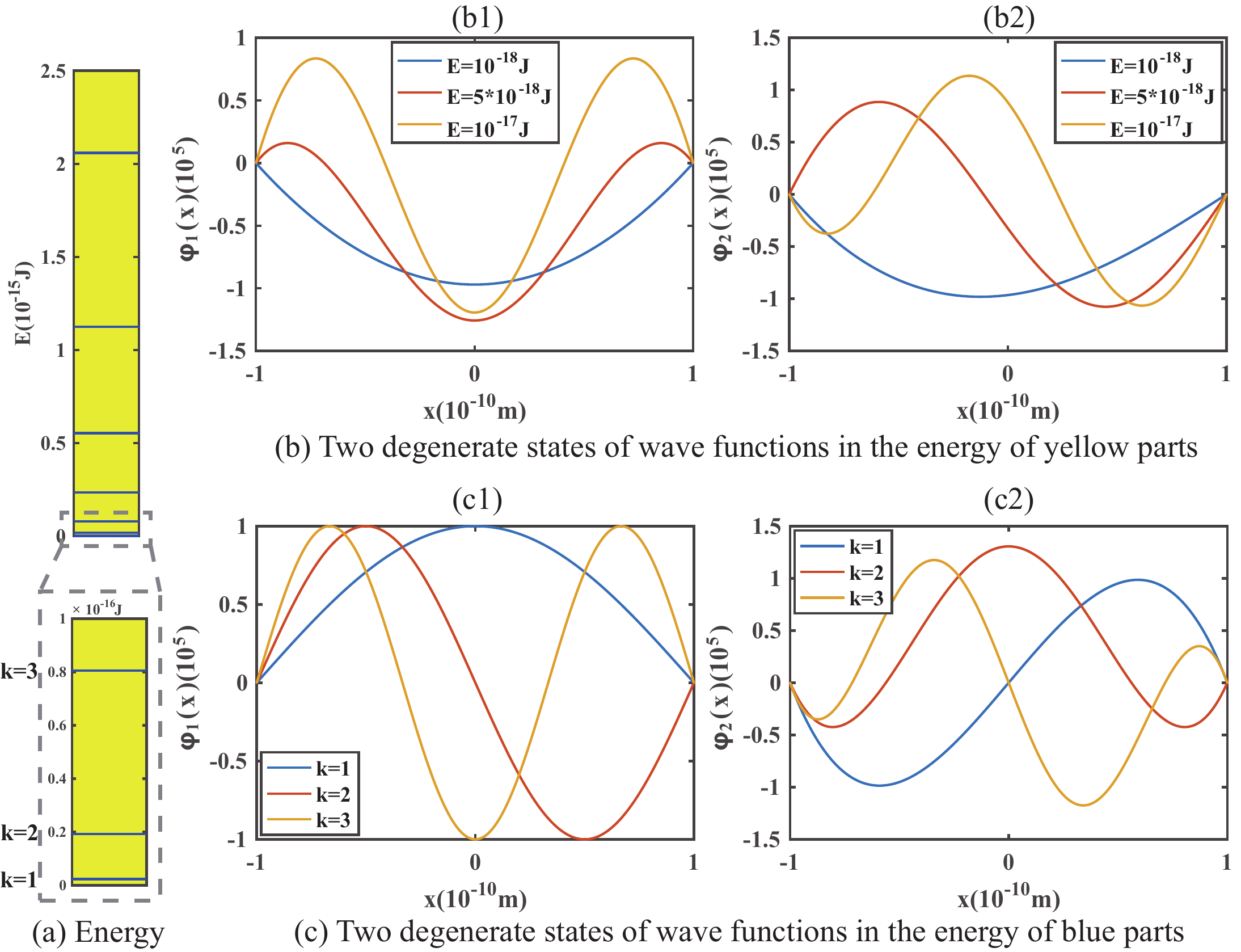}
\caption[Caption for LOF]{The wave functions and energy of infinite potential well with minimal length. Figure (a) illustrates the energy of the particle. Figure (b) illustrates wave functions in yellow energy parts of Fig. (a) for three energy $E=10^{-18}$ J, $5*10^{-18}$ J and $10^{-17}$ J with different colors, which are $2$-degree degenerate. The doubly degenerate wave functions $\varphi_1$ and $\varphi_2$ are illustrated in Figs. (b1) and (b2), respectively. Figure (c) illustrates the wave functions in the blue energy parts of Fig. (a) for $k=1,2,3$ with different colors, which are $2$-degree degenerate. The doubly degenerate wave functions $\varphi_1$ and $\varphi_2$ are illustrated in Figs. (c1) and (c2), respectively. For obvious BIC phenomena, $\beta$ here is taken to be $10^{47}$, which is still within the allowable upper bound \cite{22,23}, to make the size of $\hbar\beta[(\Delta P)^2+\langle P\rangle ^2]$ close to $\hbar$ in this potential. We choose the mass of the electron, and $a=10^{-10}$ m.}\label{fig:2}
\end{figure}

More precisely, to show the motion of particles with continuous energies, we provide the exact wave functions of each $E$ in the following.
\begin{enumerate}[(i)]\setlength{\itemsep}{0pt}
  \item\label{it:I-1} At the energy $E=k^4\pi^4\hbar^4\beta'/(16ma^4)+k^2\pi^2\hbar^2/(8ma^2)$ (blue parts in Fig. \ref{fig:2}(a)), where $k=1,2,\cdots$, the system is $2$-degree degenerate. When $\beta\rightarrow0$, these energies are equal to the standard discrete energy levels of the infinite potential well without a minimal length. For each $E$, the two wave functions are

  \begin{align*}
\varphi_1(x)&=\frac{1}{\sqrt{a}}\sin[\frac{k\pi}{2a}(x+a)];\\
\varphi_2(x)&=D_1\exp(\lambda_1 x)+D_2\exp(\lambda_2 x)+D_3\cos[\frac{k\pi}{2a}(x+a)],
  \end{align*}
  where we let $w_1(x)=\exp(\lambda_1 x)$, $w_2(x)=\exp(\lambda_2 x)$, and $w_3(x)=\cos[\frac{k\pi}{2a}(x+a)]$, then
\begin{align}\label{eq:1,2,3}
& \left\{
  \begin{array}{ll}
    D_1=\widetilde{D_1}s  \\
    D_2=\widetilde{D_2}s  \\
    D_3=\widetilde{D_3}s
  \end{array}
\right.; &
\left\{
  \begin{array}{ll}
    \widetilde{D_1}=w_3(a)w_2(-a)-w_3(-a)w_2(a)  \\
    \widetilde{D_2}=w_3(-a)w_1(a)-w_3(a)w_1(-a)  \\
    \widetilde{D_3}=w_2(a)w_1(-a)-w_2(-a)w_1(a)
  \end{array}
\right.,
\end{align}
and the parameter $s$ is determined by the equation below:
\begin{equation}\label{eq:s}
\begin{split}
s  & =\pm(\widetilde{D_1}^2F_{1,1}+\widetilde{D_2}^2F_{2,2}+\widetilde{D_3}^2F_{3,3}+
\widetilde{D_1}\widetilde{D_2}F_{1,2}+\widetilde{D_2}\widetilde{D_1}F_{2,1}\\
   & +
\widetilde{D_2}\widetilde{D_3}F_{2,3}+\widetilde{D_3}\widetilde{D_2}F_{3,2}+
\widetilde{D_1}\widetilde{D_3}F_{1,3}+\widetilde{D_3}\widetilde{D_1}F_{3,1})^{-0.5},
\end{split}
\end{equation}
where $F_{i,j}=\int_{-a}^aw_i(x)w_j^*(x)\mathrm{d}x$. Figures \ref{fig:2}(c1) and (c2) illustrate the $2$-degree degenerate states of wave functions for $k=1,2,3$. \par
  \item\label{it:I-2} At the other energy $E$ (yellow parts in Fig. \ref{fig:2}(a)), the states are also $2$-degree degenerate. These energies are extra energy under the effect of the minimal length when $\beta\rightarrow0$, which is equal to the energy gaps between discrete energy levels of the infinite potential well without the minimal length. For each $E$, the two wave functions are
    \begin{align}
  \varphi_1(x)&=D_1\exp(\lambda_1 x)+D_2\exp(\lambda_2 x)+D_3\cos(\lambda_3 x);\\
   \varphi_2(x)&=D_4\exp(\lambda_2 x)+D_5\cos(\lambda_3 x)+D_6\sin(\lambda_3 x),
  \end{align}
  where the two coefficient sets $D_1,D_2,D_3$ and $D_4,D_5,D_6$ are decided by the same method as the process Eqs. \eqref{eq:1,2,3}-\eqref{eq:s}. Figures \ref{fig:2}(b1) and (b2) illustrate the $2$-degree degenerate states of the wave functions for $E=10^{-18}$ J, $5*10^{-18}$ J and $10^{-17}$ J.
\end{enumerate}\par
We can see that the wave functions in Figs. \ref{fig:2}(b1), (b2) and (c2) are extra states caused by the perturbation of the minimal length, and the wave functions in Fig. \ref{fig:2}(c1) are the same as the standard case ($\beta'=0$). Compared to the particle that can only move with the energy in the blue parts without the minimal length, the effects of the minimal length make the particle move with the energy in the yellow parts to make the energy continuous. In particular, the quantity $k^4\pi^4\hbar^4\beta'/(16ma^4)$ of $E$ in (\ref{it:I-1}) is the energy shift, which is discussed in many papers about the minimal length \cite{11,20,21,48,49}. \par
For the environmental variables of the infinite potential well set in tpyical theoretical and experimental studies, namely, the width of the potential well $a=10^{-10}$ m and the mass of the particle $m=9.10956¡Á10^{-31}$ kg, we can calculate that when the magnitude of $\beta$ is close to $47$, which is still within the allowable upper bound of $\beta$ \cite{22,23}, $\hbar\beta[(\Delta P)^2+\langle P\rangle ^2]$ is close to the Planck constant $\hbar$. By the method in Sec. \ref{sec:C}, the BICs are obvious. \par
Although $\beta$ has a large upper bound \cite{11,22,23}, $\beta$ is often considered a small number in some researches \cite{20,21,48,49,51,52,50}, which means that $\beta[(\Delta P)^2+\langle P\rangle ^2]\ll 1$. Based on the method in Sec. \ref{sec:C}, the influence of the minimal length is negligible in these researches, and the system is close to the state with $2$ DFs (without the minimal length). Thus, under the environmental variables of the infinite potential well set as usual, the BICs exist but are very inconspicuous in the infinite potential well. This explains why BICs are not found in the infinite potential well with a single particle when the minimal length is not considered in the previous researches. On the other hand, this method in Secs. \ref{sec:B-U} and \ref{sec:C} suggests that we can adjust the environmental variables of the infinite potential well to obtain obvious BICs in the infinite potential well. \par

\subsection{Linear potential}\label{LP}
In this example, we assume a particle moving in a potential $V(x)$ satisfying $V(x)=Lx$ in the region $(0,+\infty)$ and $V(x)=+\infty$ in the other region, where $L>0$. Let $\eta^4=\beta'^{-1}$, $a=\eta^2/(4\hbar^2)$, and $b(x)=(Lx-E)m/\hbar^4$. In the region $(0,+\infty)$, the time-independent Schr\"{o}dinger equation with the minimal length becomes
\begin{equation}\label{eq:S-minimal-L}
\frac{\mathrm{d}^4\varphi(x)}{\mathrm{d}x^4}-2\eta^2a\frac{\mathrm{d}^2\varphi(x)}{\mathrm{d}x^2}+b(x)\eta^4\varphi(x)=0,
\end{equation}
where the boundary conditions are $\varphi(0)=0$ and $\lim\limits_{x\rightarrow+\infty}\varphi(x)=0$. If the energy $E>0$, solving the above Eq. \eqref{eq:S-minimal-L}, we have the approximate solutions
\begin{equation}\label{eq:S-minimal-L:s}
\varphi(x)=C_1\omega_1(x)+C_2\omega_2(x)+C_3\omega_3(x)+C_4\omega_4(x),
\end{equation}
where $C_1,\cdots,C_4$ are arbitrary coefficients, and
\begin{equation}\label{eq:omega-L}
\begin{split}
\omega_j(x)= & \frac{1}{\sqrt{\lambda_j(x)}\sqrt[4]{a^2-b(x)}}\cdot\\
&\exp[\eta\int_{x_0}^x \lambda_j(\chi)\mathrm{d}\chi-\frac{1}{2}\int_{x_0}^x\lambda_j'(\chi)\frac{1}{\sqrt{a^2-b(\chi)}}\mathrm{d}\chi][1+O(\eta^{-1})];j=1,2,3,4,
\end{split}
\end{equation}
where
\begin{align}\label{eq:lambda-L-1234}
\lambda_{1,2}(x)&=\pm\sqrt{a+\sqrt{a^2-b(x)}};& 
\lambda_{3,4}(x)&=\pm\sqrt{a-\sqrt{a^2-b(x)}}. 
\end{align}\par
We study the asymptotic behavior of $\omega_j(x)$. When $x\rightarrow+\infty$, we have
\begin{align*}
\sqrt{\lambda_j(x)}\sqrt[4]{a^2-b(x)} & \propto x^{\frac{1}{2}}; & \frac{1}{2}\int_{x_0}^x\lambda_j'(\chi)\frac{1}{\sqrt{a^2-b(\chi)}}\mathrm{d}\chi & \propto 0;
\end{align*}
where $j=1,2,3,4$, and
\begin{align*}
\exp[\int_{x_0}^x \lambda_1(\chi)\mathrm{d}\chi]\propto & \exp x^{\frac{5}{4}};& \exp[\int_{x_0}^x \lambda_2(\chi)\mathrm{d}\chi]\propto & \exp x^{-\frac{5}{4}};\\
\exp[\int_{x_0}^x \lambda_3(\chi)\mathrm{d}\chi]\propto & \exp x^{\frac{5}{4}};& \exp[\int_{x_0}^x \lambda_4(\chi)\mathrm{d}\chi]\propto & \exp x^{-\frac{5}{4}}.
\end{align*}
Thus, when $x\rightarrow+\infty$, $\omega_1,\omega_3\rightarrow+\infty$ (omitting them due to physical significance) and $\omega_2,\omega_4\rightarrow0$. The wave functions Eq. \eqref{eq:S-minimal-L:s} become
\begin{equation}\label{eq:S-minimal-L:sr}
\varphi(x)=C_2\omega_2(x)+C_4\omega_4(x).
\end{equation}\par
Substituting another boundary condition $\varphi(0)=0$ into Eq. \eqref{eq:S-minimal-L:sr}, we obtain the wave functions for any $E>0$:
\begin{equation}\label{eq:linear-w}
\varphi(x)=C_2\omega_2(x)-\frac{C_2\omega_2(0)}{\omega_4(0)}\omega_4(x),
\end{equation}
where $C_2$ can be determined after $\varphi(x)$ be normalized. These are non-degenerate since the condition $\varphi(0)=0$ reduces the DFs by $1$. By the results of case II in Sec. \ref{sec:B-U}, the energy $E$ is still continuous because the DFs remain more than zero. Thus, for any $E>0$ with continuous energy, the bound wave function is Eq. \eqref{eq:linear-w}. This indicates that the BICs exist in the potential under the effect of the minimal length.\par
For the environmental variables of the linear potential set in typical theoretical and experimental studies, namely, the linear parameter of the potential $l=mg$ and the mass of the particle $m=9.10956¡Á10^{-31}$ kg, we can calculate that when the magnitude of $\beta$ is close to $37$, which is still within the allowable upper bound of $\beta$ \cite{22,23}, $\hbar\beta[(\Delta P)^2+\langle P\rangle ^2]$ is close to the Planck constant $\hbar$. By the method in Sec. \ref{sec:C}, the BICs are obvious. \par
Although $\beta$ has a large upper bound \cite{11,22,23}, $\beta$ is often considered a small number in some researches \cite{20,21,48,49,51,52,50}, which makes $\beta[(\Delta P)^2+\langle P\rangle ^2]\ll 1$. By the method in Sec. \ref{sec:C}, the influence of the minimal length is negligible, and the system is close to the state with $2$ DFs (without the minimal length). Thus, under the environmental variables of the linear potential typically used, the BICs exist but are very inconspicuous in the linear potential. This explains why BICs are not found in the linear potential with a single particle when the minimal length is not considered in the previous researches. On the other hand, the method in Secs. \ref{sec:B-U} and \ref{sec:C} suggests that we can adjust the environmental variables of the linear potential to obtain obvious BICs in the linear potential. \par
In Ref. \cite{20,48,49}, the authors studied this potential with the minimal length, ignoring the DFs of the energy; thus, they did not find that the energy was continuous.\par

\subsection{Quantum harmonic oscillator}\label{QHO}
In this example, we assume a particle moving in the potential $V(x)$ satisfying $V(x)=0.5m\omega^2x^2$, where $\omega$ is the vibrational frequency. Let $\eta^4=\beta'^{-1}$, $a=\eta^2/(4\hbar^2)$, and $b(x)=(0.5m\omega^2x^2-E)m/\hbar^4$; the time-independent Schr\"{o}dinger equation with the minimal length becomes
\begin{equation}\label{eq:S-minimal-H}
\frac{\mathrm{d}^4\varphi(x)}{\mathrm{d}x^4}-2\eta^2a\frac{\mathrm{d}^2\varphi(x)}{\mathrm{d}x^2}+b(x)\eta^4\varphi(x)=0,
\end{equation}
where the boundary conditions are $\lim\limits_{x\rightarrow\pm\infty}\varphi(x)=0$. If the energy $E>0$, solving the above Eq. \eqref{eq:S-minimal-H}, we have the approximate solutions
\begin{equation}\label{eq:S-minimal-H:s}
\varphi(x)=C_1\omega_1(x)+C_2\omega_2(x)+C_3\omega_3(x)+C_4\omega_4(x),
\end{equation}
where $C_1,\cdots,C_4$ are arbitrary coefficients, and
\begin{equation}\label{eq:omega-H}
\begin{split}
\omega_j(x)= & \frac{1}{\sqrt{\lambda_j(x)}\sqrt[4]{a^2-b(x)}}\cdot\\
&\exp[\eta\int_{x_0}^x \lambda_j(\chi)\mathrm{d}\chi-\frac{1}{2}\int_{x_0}^x\lambda_j'(\chi)\frac{1}{\sqrt{a^2-b(\chi)}}\mathrm{d}\chi][1+O(\eta^{-1})];j=1,2,3,4,
\end{split}
\end{equation}
where
\begin{align}\label{eq:lambda-H-1234}
\lambda_{1,2}(x)&=\pm\sqrt{a+\sqrt{a^2-b(x)}};& 
\lambda_{3,4}(x)&=\pm\sqrt{a-\sqrt{a^2-b(x)}}. 
\end{align}\par
We study the asymptotic behavior of $\omega_j(x)$ when $x\rightarrow+\infty$. Similar to the above section, we have $\omega_1,\omega_3\rightarrow+\infty$ (omitting them due to physical significance) and $\omega_2,\omega_4\rightarrow0$ when $x\rightarrow\pm\infty$. Thus, the wave functions for any $E>0$ with continuous energy are
\begin{equation}\label{eq:S-minimal-H:sr}
\varphi(x)=C_2\omega_2(x)+C_4\omega_4(x),
\end{equation}
where the coefficients $C_2$ and $C_4$ cannot be determined after $\varphi(x)$ is normalized. This causes the energy to be $2$-degree degenerate. The two degenerate wave functions are $\varphi_1(x)=\omega_2(x)$ and $\varphi_2(x)=\omega_4(x)$. For any $E>0$ with continuous energy, we have obtained the bound wave functions. This indicates that the BICs exist in the potential under the effect of the minimal length.\par
For the environmental variables of the quantum harmonic oscillator set as in typical theoretical and experimental studies, namely, the vibrational frequency $\omega=10^{30}$ Hz and the mass of the particle $m=9.10956¡Á10^{-31}$ kg, we can calculate that when the magnitude of $\beta$ is close to $33$, which is still within the allowable upper bound of $\beta$ \cite{22,23}, $\hbar\beta[(\Delta P)^2+\langle P\rangle ^2]$ is close to Planck constant $\hbar$. By the method in Sec. \ref{sec:C}, the BICs are obvious. \par
Although $\beta$ has a large upper bound \cite{11,22,23}, $\beta$ is often considered a small number in some researches \cite{20,21,48,49,51,52,50}, which makes $\beta[(\Delta P)^2+\langle P\rangle ^2]\ll 1$. By the method in Sec. \ref{sec:C}, the influence of the minimal length is negligible in these researches, and the system is close to the state with $2$ DFs (without the minimal length). Thus, under the environmental variables of the quantum harmonic oscillator typical used, the BICs exist but are very inconspicuous. This explains why BICs are not found in the quantum harmonic oscillator with a single particle when the minimal length is not considered in the previous researches. On the other hand, the method in Secs. \ref{sec:B-U} and \ref{sec:C} suggests that we can adjust the environmental variables of the quantum harmonic oscillator to obtain obvious BICs. \par
In Ref. \cite{21}, the authors studied the same potential as the minimal length; they followed the method of quantum mechanics without the minimal length, and they ignored the DFs and degeneracy of the energy; thus, they did not find that the energy was continuous.\par
\section{Conclusion}
We suggest a method to analyze the spatial structure of particle bound states with a minimal length. Using this approach, we found that the BICs are universal under the minimal length. We also provided conditions for determining whether BICs are easily found. Then, we revealed a mechanism of the BICs. We think that this approach can be applied to find BICs in potentials and provide a guide to realize BICs in experiments: since the BICs are obvious when $\hbar\beta[(\Delta P)^2+\langle P\rangle ^2]$ is close to $\hbar$, we can adjust the environmental variables of the potentials to satisfy $\hbar\beta[(\Delta P)^2+\langle P\rangle ^2]$ close to $\hbar$ to obtain obvious BICs.\par
In addition, our results indicate that the minimal length, a object predicted in string theory and quantum gravity, has a definite filling effect on the energy gaps in quantum physics, which may indicate some characteristics of string theory and essential features of quantum gravity.\par

\appendix
\section{Approximations to the fourth-order Schr\"{o}dinger equation}\label{app}

Let's consider the Schr\"{o}dinger equation with the minimal length
\begin{equation*}
\frac{\beta'\hbar^4}{m}\frac{\mathrm{d}^4 \varphi(x)}{\mathrm{d} x^4}-\frac{\hbar^2}{2m}\frac{\mathrm{d}^2 \varphi(x)}{\mathrm{d} x^2}+[V(x)-E]\varphi(x)=0,
\end{equation*}\par
1. One kind of the common methods for this equation is omitting the items with $\beta'^2$ to reduce the order of the fourth-order Schr\"{o}dinger equation after using some substitution of wave functions \cite{50,54}. Since there exists the parameter $\beta'$ in the highest order item $[\beta\hbar^4\mathrm{d}^4 \varphi(x)]/(m\mathrm{d} x^4)$, the equation is a singular perturbation system \cite{55,56}, where the fourth-order term accounts for the quantum-gravitational fluctuations of the background metric \cite{53}. If we omit the term with $\beta'^2$ to reduce the order of the fourth-order Schr\"{o}dinger equation to a second-order one, the truncation causes the reduction of the dimension of the solutions space and the DFs of its solutions, thus only part of particles' states can possibly be obtained by studying the simplified second-order equation.\par
2. Another common method is studying the fourth-order Schr\"{o}dinger equation in the momentum representation. Although this method is not universal, it does work for some special potentials, such as linear potential and quantum harmonic oscillator, where the fourth-order equation can be transformed into a lower-order equation in the momentum representation \cite{12,57}. However, we should be very careful with this method since it would influence the solution space of the equation and hence alter the properties of the solution states. We will show its influence on the solutions in two different situations below. \par
The essence of this method of momentum representation is to use Fourier transform to transform the Schr\"{o}dinger equation from the $x$ representation to the $p$ representation to solve the equation. The method of the Fourier transform is a method to be used to find a hypothetical solution\footnote{``Fourier transforms may be used to find a hypothetical solution which must be verified by other means. This verification is necessary, because when the Fourier transform is applied, one is already assuming not only that a solution exists, but that it has all of the properties which are needed in order to apply the Fourier transforms, such as solution decays rapidly enough. The Fourier transform methods simply provide us with a hypothetical solution."---Ref. \cite{58}}, namely, special solutions of equation, rather than general solution of equation. The method of momentum representation is effective to solve the standard Schr\"{o}dinger equation because the solution space of the standard Schr\"{o}dinger equation has a low dimension of $2$. For example, for linear potential, its boundary conditions can restrict the extra DFs of the solutions to $0$, so there is only one wave function in each discrete energy, that is, the energy is discrete and no degeneracy. The wave function corresponds one-to-one with the hypothetical solution solved by the equation in the momentum representation, i.e., Fourier space \cite{58}.\par
However, we can not simply apply this Fourier transform method to solve the Sch\"{o}rdinger equation in the presence of the minimal length. This is because applying the Fourier transform indicates an implicit assumption that the solution space of the equation is within the domain of Fourier transform. This assumption is not valid for the Schr\"{o}dinger equation in the presence of the minimal length, and thus the Fourier transform method will artificially reduce the order of the solution space (and the degree of freedom) of the equation when converting the equation in the momentum representation. In fact, we have showed that the boundary conditions of the linear potential and the quantum harmonic oscillator are not enough to reduce the extra DFs to $0$ in the presence of the minimal length, which implies that the energy levels are continuous or degenerate in the presence of the minimal length (see details in Sec. \ref{LP} and \ref{QHO}). However, the extra DFs of the equation are artificially forced to zero after transformed in the momentum representation. Therefore, the hypothetical solution in the momentum representation does not correspond one-to-one with the wave function of the original equation. In other words, the hypothetical solution is only a certain solution of the original equation, rather than the general solution, namely, not all the states of particles are found. \par
More precisely, for example for the linear potential, in the position representation, we have proved the dimension of the solution space of the Schr\"{o}dinger equation in the presence of the minimal length is $4$ (see more details in Sec. \ref{sec:B-U}). In the momentum representation, the Schrodinger equation is reduced to
\begin{equation*}
(i\hbar l+i\hbar\beta lp^2)\frac{\partial}{\partial p} C(p)+(\frac{p^2}{2m}-E)C(p)=0.
\end{equation*}
\begin{figure}[h]
\centering
\includegraphics[width=11cm]{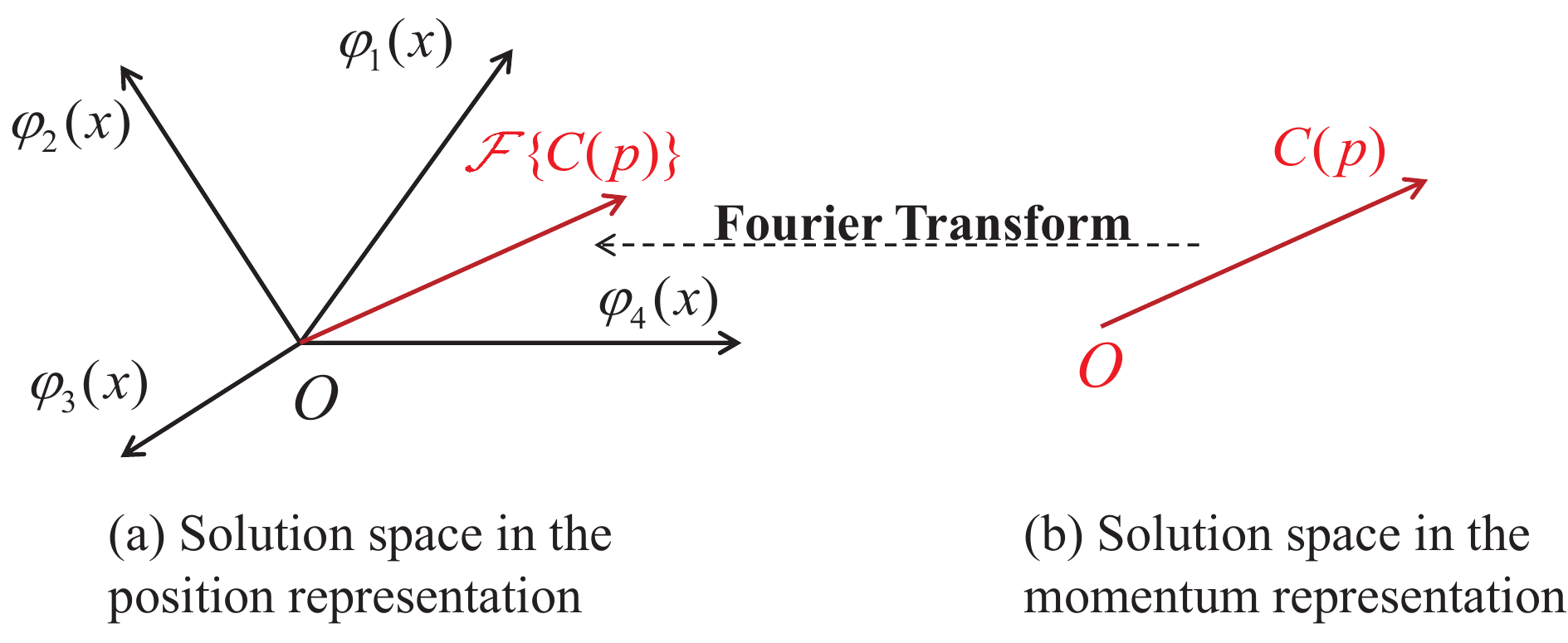}
\caption{The solution space of the Schr\"{o}dinger equation in the presence of the minimal length for the linear potential. $\varphi_1(x),\varphi_2(x),\varphi_3(x),\varphi_4(x)$ is a set of basis in the solution space in the position representation, and $\mathcal{F}\{C(p)\}$ is the Fourier transform of $C(p)$, where $C(p)$ is a set of basis in the solution space in the momentum representation.}\label{fig:4}
\end{figure}
Obviously, the dimension of the solution space of the Schr\"{o}dinger equation in the presence of the minimal length is $1$. Actually, only particular solution can be obtained by solving the equation in the momentum representation, as shown in Fig. \ref{fig:4} \cite{58,59}.

\par

\acknowledgments
The work is supported by National Natural Science Foundation of China for the Young (No. 11801385) and China Postdoctoral Science Foundation (No. 2017M620425).
\bibliographystyle{JHEP}
\bibliography{beginning6reference0}

\providecommand{\href}[2]{#2}\begingroup\raggedright\begin{thebibliography}{10}

\bibitem{1}
J.~von Neumann and E.~Wigner, \emph{\"{U}ber merkw\"{u}rdige diskrete
  eigenwerte}, {\emph{Phys. Z.(in German)} {\bfseries 30} (1929) 465}.

\bibitem{2}
F.~H. Stillinger and D.~R. Herrick, \emph{Bound states in the continuum},
  \href{https://doi.org/10.1103/PhysRevA.11.446}{\emph{Phys. Rev. A} {\bfseries
  11} (1975) 446}.

\bibitem{24}
A.~K. Jain and C.~S. Shastry, \emph{Bound states in the continuum for separable
  nonlocal potentials},
  \href{https://doi.org/10.1103/PhysRevA.12.2237}{\emph{Phys. Rev. A}
  {\bfseries 12} (1975) 2237}.

\bibitem{8}
N.~Moiseyev, \emph{Suppression of feshbach resonance widths in two-dimensional
  waveguides and quantum dots: A lower bound for the number of bound states in
  the continuum},
  \href{https://doi.org/10.1103/PhysRevLett.102.167404}{\emph{Phys. Rev. Lett.}
  {\bfseries 102} (2009) 167404}.

\bibitem{7}
Y.~Plotnik, O.~Peleg, F.~Dreisow, M.~Heinrich, S.~Nolte, A.~Szameit et~al.,
  \emph{Experimental observation of optical bound states in the continuum},
  \href{https://doi.org/10.1103/PhysRevLett.107.183901}{\emph{Phys. Rev. Lett.}
  {\bfseries 107} (2011) 183901}.

\bibitem{3}
Z.~Xiao, W.~Chaozhen, L.~Yingming and L.~Maokang, \emph{Phase transitions of
  energy and wave functions and bound states in the continuum},
  \href{https://doi.org/10.1103/PhysRevA.93.042106}{\emph{Phys. Rev. A}
  {\bfseries 93} (2016) 042106}.

\bibitem{4}
Z.~Xiao, Y.~Bo, W.~Chaozhen and L.~Maokang, \emph{The transition of energy and
  bound states in the continuum of fractional schr\"{o}dinger equation in
  gravitational field and the effect of the minimal length},
  \href{https://doi.org/10.1016/j.cnsns.2018.05.002}{\emph{Commun. Nonlinear
  Sci.} {\bfseries 67} (2019) 290}.

\bibitem{6}
F.~Capasso, C.~Sirtori, J.~Faist, D.~L. Sivco, S.-N.~G. Chu and A.~Y. Cho,
  \emph{Observation of an electronic bound state above a potential well},
  \href{https://doi.org/10.1038/358565a0}{\emph{Nature} {\bfseries 358} (1992)
  565}.

\bibitem{10}
D.~C. Marinica, A.~G. Borisov and S.~Shabanov, \emph{Bound states in the
  continuum in photonics},
  \href{https://doi.org/10.1103/PhysRevLett.100.183902}{\emph{Phys. Rev. Lett.}
  {\bfseries 100} (2008) 183902}.

\bibitem{25}
L.~Yuan and Y.~Y. Lu, \emph{Bound states in the continuum on periodic
  structures surrounded by strong resonances},
  \href{https://doi.org/10.1103/PhysRevA.97.043828}{\emph{Phys. Rev. A}
  {\bfseries 97} (2018) 043828}.

\bibitem{26}
P.~Facchi, D.~Lonigro, S.~Pascazio, F.~V. Pepe and D.~Pomarico, \emph{Bound
  states in the continuum for an array of quantum emitters},
  \href{https://doi.org/10.1103/PhysRevA.100.023834}{\emph{Phys. Rev. A}
  {\bfseries 100} (2019) 023834}.

\bibitem{27}
A.~Khelashvili and N.~Kiknadze, \emph{Bound states in continuum induced by
  relativity}, \href{https://doi.org/10.1103/PhysRevA.55.2557}{\emph{Phys. Rev.
  A} {\bfseries 55} (1997) 2557}.

\bibitem{28}
M.~Ahumada, P.~A. Orellana and J.~C. Retamal, \emph{Bound states in the
  continuum in whispering gallery resonators},
  \href{https://doi.org/10.1103/PhysRevA.98.023827}{\emph{Phys. Rev. A}
  {\bfseries 98} (2018) 023827}.

\bibitem{5}
C.~W. Hsu, B.~Zhen, J.~Lee, S.-L. Chua, S.~G. Johnson, J.~D. Joannopoulos
  et~al., \emph{Observation of trapped light within the radiation continuum},
  \href{https://doi.org/10.1038/nature12289}{\emph{Nature} {\bfseries 499}
  (2013) 188}.

\bibitem{9}
B.~Zhen, C.~W. Hsu, L.~Lu, A.~D. Stone and M.~Solja\v{c}i\'{c},
  \emph{Topological nature of optical bound states in the continuum},
  \href{https://doi.org/10.1103/PhysRevLett.113.257401}{\emph{Phys. Rev. Lett.}
  {\bfseries 113} (2014) 257401}.

\bibitem{18}
A.~Albo, D.~Fekete and G.~Bahir, \emph{Electronic bound states in the continuum
  above (ga,in)(as,n)/(al,ga)as quantum wells},
  \href{https://doi.org/10.1103/PhysRevB.85.115307}{\emph{Phys. Rev. B}
  {\bfseries 85} (2012) 115307}.

\bibitem{15}
L.~D. Landau and E.~M. Lifshits, \emph{Quantum Mechanics (Non-Relativistic
  Theory)}. Higher Education Press, Beijing, 6th~ed., 2008.

\bibitem{30}
Y.~Boretz, G.~Ordonez, S.~Tanaka and T.~Petrosky, \emph{Optically tunable bound
  states in the continuum},
  \href{https://doi.org/10.1103/PhysRevA.90.023853}{\emph{Phys. Rev. A}
  {\bfseries 90} (2014) 023853}.

\bibitem{65}
J.~Mur-Petit and R.~A. Molina, \emph{Chiral bound states in the continuum},
  \href{https://doi.org/10.1103/PhysRevB.90.035434}{\emph{Phys. Rev. B}
  {\bfseries 90} (2014) 035434}.

\bibitem{66}
J.~Mur-Petit and R.~A. Molina, \emph{Van hove bound states in the continuum:
  localised subradiant states in finite open lattices},
  \href{https://doi.org/arXiv:2002.05959v1}{\emph{arXiv} (2020) }.

\bibitem{11}
S.~Das and E.~C. Vagenas, \emph{Universality of quantum gravity corrections},
  \href{https://doi.org/10.1103/PhysRevLett.101.221301}{\emph{Phys. Rev. Lett.}
  {\bfseries 101} (2008) 221301}.

\bibitem{12}
A.~Kempf, G.~Mangano and R.~B. Mann, \emph{Hilbert space representation of the
  minimal length uncertainty relation},
  \href{https://doi.org/10.1103/PhysRevD.52.1108}{\emph{Phys. Rev. D}
  {\bfseries 52} (1995) 1108}.

\bibitem{20}
F.~Brau and F.~Buisseret, \emph{Minimal length uncertainty relation and
  gravitational quantum well},
  \href{https://doi.org/10.1103/PhysRevD.74.036002}{\emph{Phys. Rev. D}
  {\bfseries 74} (2006) 036002}.

\bibitem{21}
I.~Dadic, L.~Jonke and S.~Meljanac, \emph{Harmonic oscillator with minimal
  length uncertainty relations and ladder operators},
  \href{https://doi.org/10.1103/PhysRevD.67.087701}{\emph{Phys. Rev. D}
  {\bfseries 67} (2003) 087701}.

\bibitem{22}
G.~Lambiase and F.~Scardigli, \emph{Lorentz violation and generalized
  uncertainty principle},
  \href{https://doi.org/10.1103/PhysRevD.97.075003}{\emph{Phys. Rev. D}
  {\bfseries 97} (2018) 075003}.

\bibitem{23}
F.~Scardigli and R.~Casadio, \emph{Gravitational tests of the generalized
  uncertainty principle},
  \href{https://doi.org/10.1140/epjc/s10052-015-3635-y}{\emph{Eur. Phys. J. C}
  {\bfseries 75} (2015) 425}.

\bibitem{32}
S.~Jalalzadeh, S.~Rasouli and P.~Moniz, \emph{Quantum cosmology, minimal
  length, and holography},
  \href{https://doi.org/10.1103/PhysRevD.90.023541}{\emph{Phys. Rev. D}
  {\bfseries 90} (2014) 023541}.

\bibitem{35}
J.~Y. Bang and M.~S. Berger, \emph{Quantum mechanics and the generalized
  uncertainty principle},
  \href{https://doi.org/10.1103/PhysRevD.74.125012}{\emph{Phys. Rev. D}
  {\bfseries 74} (2006) 125012}.

\bibitem{48}
K.~Nozari and P.~Pedram, \emph{Minimal length and bouncing-particle spectrum},
  \href{https://doi.org/10.1209/0295-5075/92/50013}{\emph{Europhys. Lett.}
  {\bfseries 92} (2010) 5}.

\bibitem{49}
L.~N. Chang, Z.~Lewis, D.~Minic and T.~Takeuchi, \emph{On the minimal length
  uncertainty relation and the foundations of string theory},
  \href{https://doi.org/10.1155/2011/493514}{\emph{Adv. High Energy Phys.}
  {\bfseries 2011} (2011) 493514}.

\bibitem{62}
A.~F. Ali, \emph{No existence of black holes at lhc due to minimal length in
  quantum gravity}, \href{https://doi.org/10.1007/JHEP09(2012)067}{\emph{J.
  High Energ. Phys.} {\bfseries 67} (2012) }.

\bibitem{63}
Y.~C. Ong, \emph{An effective black hole remnant via infinite evaporation time
  due to generalized uncertainty principle},
  \href{https://doi.org/10.1007/JHEP10(2018)195}{\emph{J. High Energ. Phys.}
  {\bfseries 195} (2018) }.

\bibitem{61}
M.~Bishop, J.~Lee and D.~Singleton, \emph{Modified commutators are not
  sufficient to determine a quantum gravity minimal length scale},
  \href{https://doi.org/10.1016/j.physletb.2020.135209}{\emph{Phys. Lett. B}
  {\bfseries 802} (2020) 135209}.

\bibitem{51}
A.~Kempf and L.~Lorenz, \emph{Exact solution of inflationary model with minimum
  length}, \href{https://doi.org/10.1103/PhysRevD.74.103517}{\emph{Phys. Rev.
  D} {\bfseries 74} (2006) 103517}.

\bibitem{52}
A.~Ashoorioon, A.~Kempf and R.~B. Mann, \emph{Minimum length cutoff in
  inflation and uniqueness of the action},
  \href{https://doi.org/10.1103/PhysRevD.71.023503}{\emph{Phys. Rev. D}
  {\bfseries 71} (2005) 023503}.

\bibitem{64}
M.~Isi, J.~Mureika and P.~Nicolini, \emph{Self-completeness and the generalized
  uncertainty principle},
  \href{https://doi.org/10.1007/JHEP11(2013)139}{\emph{J. High Energ. Phys.}
  {\bfseries 802} (2013) }.

\bibitem{13}
J.~Zeng, \emph{Quantum Mechanics}. Science Press, Beijing, 4th~ed., 2007.

\bibitem{14}
R.~Su, \emph{Quantum Mechanics}. Higher Education Press, Beijing, 2nd~ed.,
  2002.

\bibitem{50}
S.~Haouat, \emph{Schr\"{o}dinger equation and resonant scattering in the
  presence of a minimal length},
  \href{https://doi.org/10.1016/j.physletb.2013.12.060}{\emph{Phys. Lett. B}
  {\bfseries 729} (2014) 33}.

\bibitem{54}
H.~Hassanabadi, S.~Zarrinkamar and E.~Maghsoodi, \emph{Scattering states of
  woods¨csaxon interaction in minimal length quantum mechanics},
  \href{https://doi.org/10.1016/j.physletb.2012.11.005}{\emph{Phys. Rev. D}
  {\bfseries 718} (2012) 678}.

\bibitem{57}
L.~N. Chang, D.~Minic, N.~Okamura and T.~Takeuchi, \emph{Exact solution of the
  harmonic oscillator in arbitrary dimensions with minimal length uncertainty
  relations}, \href{https://doi.org/10.1103/PhysRevD.65.125027}{\emph{Phys.
  Rev. D} {\bfseries 65} (2002) 125027}.

\bibitem{55}
T.~Kato, \emph{Perturbation Theory For Linear Operator}. World Publishing
  Corporation, Beijing, 2016.

\bibitem{56}
M.~Zhou, W.~Lin, M.~Ni and et~al, \emph{Perturbation Theory For Linear
  Operator}. Science Press, Beijing, 2014.

\bibitem{53}
M.~S. Berger and M.~Maziashvili, \emph{Free particle wave function in light of
  the minimum-length deformed quantum mechanics and some of its
  phenomenological implications},
  \href{https://doi.org/10.1103/PhysRevD.84.044043}{\emph{Phys. Rev. D}
  {\bfseries 84} (2011) 044043}.

\bibitem{58}
D.~Bleecker and G.~Csordas, \emph{BASIC PARTIAL DIFFERENTIAL EQUATIONS}. Van
  Nostrand Reinhold, New York, 1st~ed., 1992.

\bibitem{59}
K.~Ryuo, \emph{Fourier Analysis}. Higher Education Press, Beijing, 1st~ed.,
  1984.

\bibitem{19}
A.~D. Polyanin and V.~F. Zaitsev, \emph{Handbook of Exact Solutions for
  Ordinary Differential Equations}. Chapman Hall/CRC,, Boca Raton, 2th~ed.,
  2003.

\bibitem{16}
M.~Belloni and R.~Robinett, \emph{The infinite well and dirac delta function
  potentials as pedagogical, mathematical and physical models in quantum
  mechanics}, \href{https://doi.org/10.1016/j.physrep.2014.02.005}{\emph{Phys.
  Rep.} {\bfseries 540} (2014) 25}.

\end{thebibliography}\endgroup
\end{document}